# Skin Tone Emoji and Sentiment on Twitter

Steven Coats[1][0000-0002-7295-3893]

[1] English Philology, University of Oulu, 90014 Oulu, Finland
`steven.coats (at) oulu.fi`

**Abstract.** In 2015, the Unicode Consortium introduced five skin tone emoji that can be used in combination with emoji representing human figures and body parts. In this study, use of the skin tone emoji is analyzed geographically in a large sample of data from Twitter. It can be shown that values for the skin tone emoji by country correspond approximately to the skin tone of the resident populations, and that a negative correlation exists between tweet sentiment and darker skin tone at the global level. In an era of large-scale migrations and continued sensitivity to questions of skin color and race, understanding how new language elements such as skin tone emoji are used can help frame our understanding of how people represent themselves and others in terms of a salient personal appearance attribute.

## 1 Introduction

### 1.1 Background

Unicode code points are used not only to map the characters of the world's languages, but since 2009 also for emoji – characters that often depict faces or human forms.[1] Introduced by Japanese telecommunications providers in the 1990s, emoji were implemented in the popular iOS and Android mobile operating systems as well as on Social Media platforms such as Facebook, Twitter, or Instagram shortly after their canonization in the Unicode scheme. In 2015 the Unicode consortium introduced a new set of emoji characters that include code points allowing users to select from five different skin tones, in addition to a default skin tone (usually yellow, Fig. 1), for a set of emoji characters that depict persons and body parts [1]. The skin tones, derived from the Fitzpatrick scale used in dermatology, are applied to a face or body-part emoji by appending the Unicode code point for the skin tone to the code point for the face or body part.

In this study the use of the skin tone emoji in a large global dataset of messages collected from Twitter is investigated. After characterizing the global distribution of skin tone emoji, a sentiment analysis is conducted. The correlation of skin tone emoji and sentiment may reflect demographic and economic realities but can also shed light on evolving attitudes towards skin color, race and ethnicity.

---

[1] A list of can be found at https://emojipedia.org.

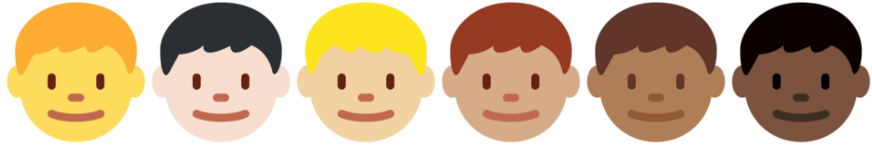

**Fig. 1.** Twitter emoji in default yellow and in the five Fitzpatrick scale-derived values

Sentiment analysis, or the automatic extraction of opinions or emotions from text data, is an important topic in Natural Language Processing. Approaches in sentiment analysis range from lexicon-based frequency counts (the "bag-of-words" model) to the use of machine learning techniques based on the automatic extraction of features in multi-dimensional vector space or the use of neural networks (for an overview, see [2]). The approach adopted in this paper utilizes an existing emoji sentiment classification scale [3] to annotate tweets with sentiment.

### 1.2 Organization of the Text

In the next section related work on emoji and skin tone emoji is described, as well as methods for sentiment analysis relevant to the present research. In Section 3, the collection and processing of a data set from the Twitter APIs and the tools and methods used to undertake the analysis are introduced. In Section 4, the results of two experiments are presented. In Section 5, the results are interpreted, a preliminary conclusion is reached, and an outlook for further investigation of skin tone emoji is offered.

## 2 Related Work

### 2.1 Work on Emoticons and Emoji in Twitter

Due to the newness of the phenomenon, analyses of skin tone emoji use are relatively few, but some research has investigated patterns of emoji usage in general. Emoticons, older ASCII-character sequences used to represent mainly facial expressions, have a longer history in Computer-mediated Communication (CMC), and have been subject to several analyses, including of their use on Twitter [4, 5, 6, 7].

For emoji, Barbieri et al. [8] used vector space representations to compare the meanings of emoji in Twitter corpora of American English, British English, peninsular Spanish and Italian. They note that while the semantics of emoji across languages and varieties are relatively stable, some emoji are used quite differently in the corpora.

McGill [9] drew attention to the underrepresentation of lighter skin-tone emoji in the United States, and suggested that while the default yellow skin tone may be used by some as a stand-in for lighter skin tones, people of European descent in the United States may also be fearful of asserting their racial identity.

Kralj-Novak et al. [3] engaged annotators to rate the sentiment of Twitter messages containing emoji in 13 languages. The derived sentiment values for individual emoji are utilized in Section 4 to assign sentiment to the data collected for this study.

Ljubešić and Fišer [10] demonstrated that Twitter users who make use of emoji tend to be more active on the platform than non-emoji users, as well as have more followers and friends. They note that the "Emoji modifier Fitzpatrick type-1-2", encoding light skin tone, is one of the most frequent emoji in their data set, comprising 2.3% of all emoji forms (85). In terms of geographic distribution, they note that clustering nations on the basis of emoji probability distributions results in a stratification of the skin tone emoji, with lighter skin tones among the most characteristic types in "first- and second-world" nations and darker skin tones more characteristic for the "fourth-world" cluster comprising mainly African nations (86–87).

## 2.2 Twitter Sentiment Analysis

Many sentiment analysis studies have utilized data from Twitter [11, 12], and sentiment analysis of monolingual labelled data can typically attain high rates of precision and accuracy. Sentiment analysis of multilingual data, on the other hand, poses various problems: For some languages there are no existing resources such as sentiment lexicons or sentiment-labelled corpora with which supervised models could be trained. Where multilingual sentiment analysis has been undertaken, it often targets specific language pairs or a small number of languages. Even if sentiment-labelled corpora exist, low levels of annotator agreement can place an upper limit on the accuracy of models [13].

Emoticons and emoji can be utilized in unsupervised sentiment analysis on the basis of the fact that they are used in many languages. Tang et al. [14], for example, used ASCII-based emoticons in Twitter messages to create a sentiment classifier using neural networks. Jiang et al. [15] used machine learning to create an "Emotion Space Model" from emoji-containing data obtained from Sina Weibo (a Chinese microblogging service similar to Twitter).

In this study, a similar approach has been adopted. Manual annotation of the tweets in the data was not undertaken, but rather sentiment values assigned on the basis of aggregate use of emoji in the Kralj-Novak et al. emoji sentiment lexicon. Examination of the labeled data suggests that the approach can offer acceptable results.

## 3 Data Collection and Processing

653,457,659 tweets with "place" metadata were collected from the Twitter Streaming API from November 2016 until June 2017 and stored at servers operated by Finland's Centre for Scientific Computing.[2] From Unicode's list of all emoji,[3] regular expressions were used to identify the 102 unique emoji types that can be used with skin tone modifiers on the Twitter platform (as of late 2017).

## 4 Analysis

In a first analysis, the prevalence of use of the skin tone emoji was determined by country and the median skin tone values calculated and mapped. Semantic properties of the skin tone emoji were investigated using vector representations, and the relationship between mean skin tone values and sentiment was considered by aggregating tweets at the level of country or territory.

### 4.1 Geographic Distribution of Skin Tone Emoji

For each of the 247 country-level administrative units in the data, frequencies of the default emoji and the skin-tone modified emoji were calculated (Table 1 summarizes the results for the 10 countries with the most tweets).

**Table 1.** Counts of tweets, potential skin tone emoji, and skin tone emoji for the 10 countries with the most tweets.

| Country | Tweets | Pot. skintone | Skin tone | Proportion |
|---|---|---|---|---|
| USA | 201,361,543 | 11,153,159 | 8,605,451 | 0.77 |
| Brazil | 92,987,119 | 3,763,962 | 1,497,697 | 0.40 |
| Japan | 38,598,876 | 2,228,136 | 445,699 | 0.20 |
| Great Britain | 35,837,868 | 2,497,332 | 1,387,807 | 0.56 |
| Philippines | 20,808,246 | 1,269,572 | 581,922 | 0.46 |
| Argentina | 20,023,675 | 1,836,721 | 290,912 | 0.16 |
| Turkey | 19,218,332 | 1,005,601 | 356,020 | 0.35 |
| Spain | 15,822,869 | 1,086,395 | 361,905 | 0.33 |
| Malaysia | 14,787,098 | 811,196 | 432,322 | 0.53 |
| France | 13,825,568 | 763,565 | 371,576 | 0.49 |

---

[2] For this data, a high rate of correlation exists between "place" latitude-longitude coordinates and "geo" latitude-longitude coordinates for tweets that contain both metadata fields. As tweets with "place" attributes are far more numerous than tweets with "geo" attributes, they are considered to be an accurate indication of user location.
[3] http://unicode.org/emoji/charts/full-emoji-list.html

Globally, more than 25 million tweets contained emoji that could take skin tone values, and these tweets contained approximately 19 million skin tone emoji. 54.3% of tweets with at least one potential skin tone emoji had an emoji with an assigned skin tone value, and 50.1% of potential skin tone emoji had skin tone values. Users in the United States, the country of origin of Twitter, of the Unicode standard, and of the skin-tone emoji, are more likely to use the skin tone modifiers. Anglophone countries such as Britain and the Philippines also use relatively many skin tone emoji. The proportion of skin-tone-possible emoji that were assigned skin tone according to country is shown in Fig. 2.

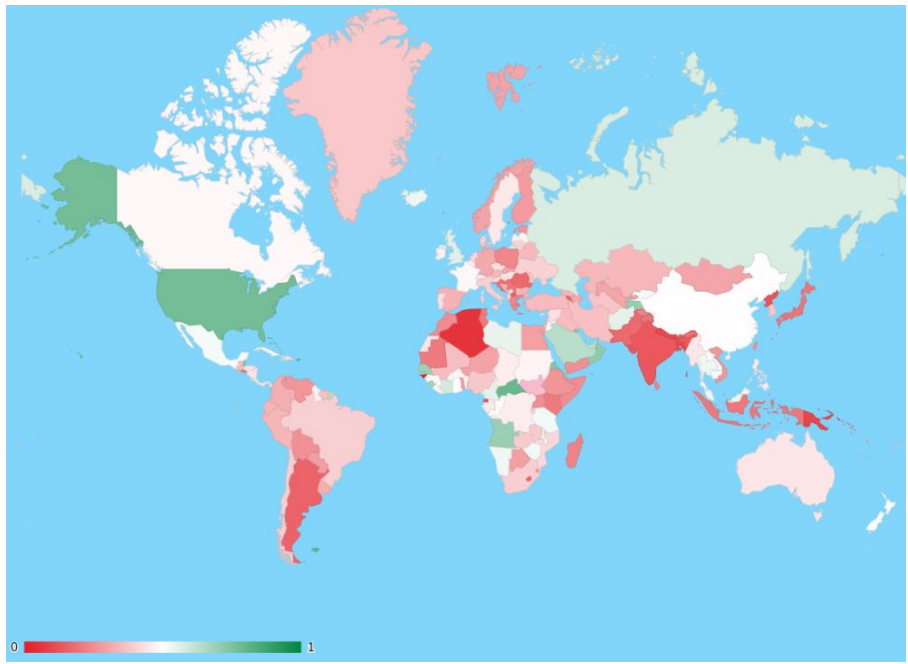

**Fig. 2.** Proportion of emoji with skin tone values. Green shades indicate a higher proportion of skin-tone emoji; red shades lower.

### 4.2 Median Skin Tone Values

The global distribution of skin tone values was as follows: light, 36%; medium-light 25%; medium 20%, medium-dark, 16%, dark, 3%. To some extent, the median skin tone value by country/territory (Fig. 3) corresponds with levels of yearly insolation, which in turn affects the average level of skin pigmentation in ancestral human population groups (Fig. 4). Lighter-than-expected skin tone values in Asian countries may reflect cultural values associating lighter skin with health and beauty. The higher value for Afghanistan may be due to the presence of U.S. military personnel in the country. Darker-than-expected skin tone values in the United States may reflect the disproportional popularity of Twitter among African-Americans (see [16]). For Europe, darker median skin tone values may reflect enthusiasm for African or African-American popular culture or migration.

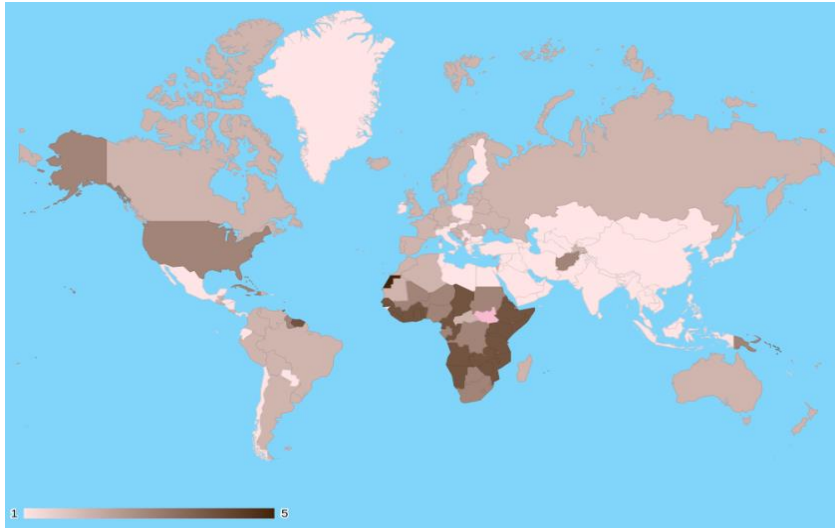

**Fig. 3.** Median value of skin tone emoji by country

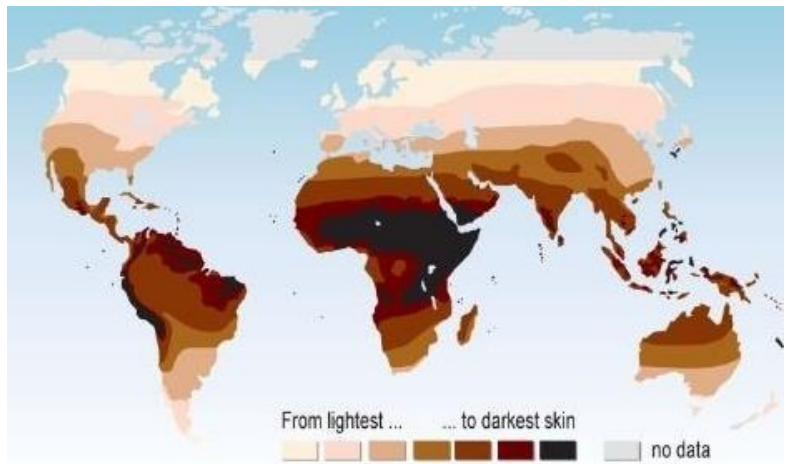

**Fig. 4.** Skin color of ancestral populations (source: [17])

**4.3   Emoji Skin Tone and Sentiment**

Two methods were used to investigate the sentiment of the corpus. In a first experiment, word embeddings in multidimensional vector space were created to identify lexical items close to the skin tone emoji in meaning. In a second experiment, sentiment per tweet was calculated by utilizing the Kralj-Novak emoji sentiment classification lexicon.

**Word Embeddings in Multidimensional Space**

Recent work in many types of Natural Language Processing has seen widespread use of word embeddings for tasks ranging from translation to content extraction, part-of-speech tagging, parsing, or sentiment analysis. The basic principle underlying these approaches was alluded to by Firth's dictum that one shall "know a word by the company it keeps" [18]. First formally proposed by Harris [19] and sometimes referred to as the "Distributional Hypothesis", it refers to the fact that linguistic elements that show similar collocational and syntactic distributions often exhibit similar semantics; measures such as pointwise mutual information can be incorporated into models that quantify the collocational properties of words or n-grams. In word embedding models, the words in a document or set of documents can be transformed into vectors based on the probability of their co-occurrence within a specified span.

The 25,297,245 tweets that contained emoji that could potentially take on skin-tone values were used to create word embeddings: All unique tokens in these tweets were assigned values in a 400-dimensional vector space based on an continuous bag-of-words embedding window of five tokens to the right and left and a minimum of 10 token occurrences in the corpus, using an implementation of the Word2Vec algorithm [20, 21]. A preliminary insight into the differences in meaning the skin tone emoji can entail is provided by examining the tokens closest to the skin tone code points in the resulting vector space (Fig. 5).

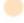

**Fig. 5.** Ten closest tokens in vector space for the five skin-tone emoji

The cosine distance indicates the similarity of the vectors for the token pair and can range in value from -1 (opposite vectors, semantics very different) to 1 (identical vectors, semantics very similar). Among the most similar tokens for all five types are the other skin tone emoji. This suggests that the semantic value of the skin tone emoji is well represented by the vector space model, and is to be expected based on the fact that these characters occur in the same contexts (as code points following a limited set of face- or body-part code points).

The types most similar to each of the skin tones in terms of cosine distance give some insight into the contexts of use of each of the skin tone emoji and hence their meanings. Light skin tone values are associated with emoji that express affection, satisfaction, or happiness. Medium-light skin tones are associated with mainly positive emoji expressing affection or irreverence. Medium skin tone emoji are closest to emoji with negative affective connotations such as crying and shouting faces, the two-eyes emoji (possibly used as an expression of incredulity), and a skull emoji, as well as a character with the numeral *100* and the positive "face with tears of joy" emoji.[4] Medium-dark skin tones are additionally associated with the informal English-language words *lol, tho* and *bruh*. The dark skin tone is closest to *100*, the two eyes, the skull, and emoji representing fire, prayer, and speaking.

---

[4] The *100* emoji was originally used in Japanese mobile communications to indicate a teacher's mark of 100 points for a school assignment, but in American usage is likely related to *keep it 100*, a phrase meaning to *keep it real*, or "be honest/authentic".

**Labelling sentiment of Tweets**

Kralj-Novak et al. [3] provide frequency information for the annotation of tweets containing emoji in 13 languages as "negative", "neutral", or "positive" by manual annotators.[5] Examples of emoji with positive and negative sentiment values are shown in Table 2: Faces expressing, for example, indifference or anger have negative values, while faces expressing affection, flowers, or gifts have positive values.[6] From this lexicon, scores for individual emoji were calculated by subtracting the number of negatively-evaluated sentences containing a particular emoji from the number of positively evaluated sentences with the same type and dividing by the total number of occurrences of that emoji.

**Table 2.** Examples of emoji with negative and positive sentiment ratings

| Emoji | sentiment | Emoji | sentiment |
|---|---|---|---|
| 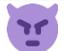 | -0.56 | 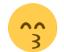 | 0.96 |
| 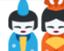 | -0.48 | 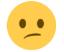 | 0.81 |
| 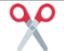 | -0.45 | 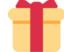 | 0.80 |
| 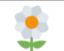 | -0.4 | 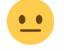 | 0.77 |
| 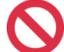 | -0.39 | 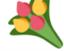 | 0.76 |

Emoji with at least 50 occurrences in the Kralj-Novak data were used to evaluate the sentiment of the 653.5 million tweets in the experimental data. Tweets were cleaned of usernames, hashtags and urls, converted to lower case, and tokenized using the NLTK Twitter Tokenizer [22], the Jieba tokenizer for Mandarin [23] and the Tiny Segmenter for Japanese [24], then scored using the Kralj-Novak sentiment scale. Examples are shown in Fig. 6.[7]

---

[5] http://kt.ijs.si/data/Emoji_sentiment_ranking/
[6] The "Japanese Dolls" emoji represents figures used in a traditional Japanese observance.
[7] Negative emoji are less frequently used in the data. Negative sentiment, in general, is less frequently expressed [25].

| | | | | |
|---|---|---|---|---|
| BR | pt | Ser rejeitado é horrível 👎 | -0.189922 | [ser, rejeitado, é, horrível, 👎] |
| JP | en | Great read @tmase04 👍 https://t.co/QrVkGZBArc | 0.522114 | [great, read, 👍] |
| US | en | Repost from @xjoshalvesx 🙏 Thank you fir yesterday! Thank you @BrianFrasierM for the pairs… https://t.co/VswnVVzZdD | 0.417804 | [repost, from, 🙏, ⚫, thank, you, fir, yesterday, !, thank, you, for, the, pairs, …] |
| AR | es | Tengo #nexflix no me interesa mas la vida! 👌❤ | 1.162317 | [tengo, no, me, interesa, mas, la, vida, !, 👌, 🟡, ❤, ] |

**Fig. 6.** Four tweets in which country of origin, automatically detected language, original text, calculated emoji-sentiment value, and tokens after cleaning are shown.

**Correlation of Skin Tone and Sentiment**

Mean skin tone per tweet was calculated by assigning the values 1 to 5 to the skin tone emoji, then for each tweet, dividing the sum of the skin tone emoji values by the number of skin tone emoji.[8] Mean skin tone and sentiment were correlated in all tweets containing at least one skin tone emoji for the entire data and at country/territorial level by using Pearson's product-moment correlation. For the entire data, a weak negative correlation between sentiment and skin tone values was found ($r$ = -0.09, $df$ = 13,736,953, $p < 10^{-32}$).

At the level of country/territory, the correlation between mean sentiment and mean skin tone value was more strongly negative, at $r$ = -0.25 ($df$ = 235, $p$ = 0.000076) for the 237 countries or territories with at least one skin tone emoji (Fig. 7).

---

[8] The values assigned were: light skin tone = 1, medium-light skin tone = 2, medium skin tone = 3, medium-dark skin tone = 4, dark skin tone = 5.

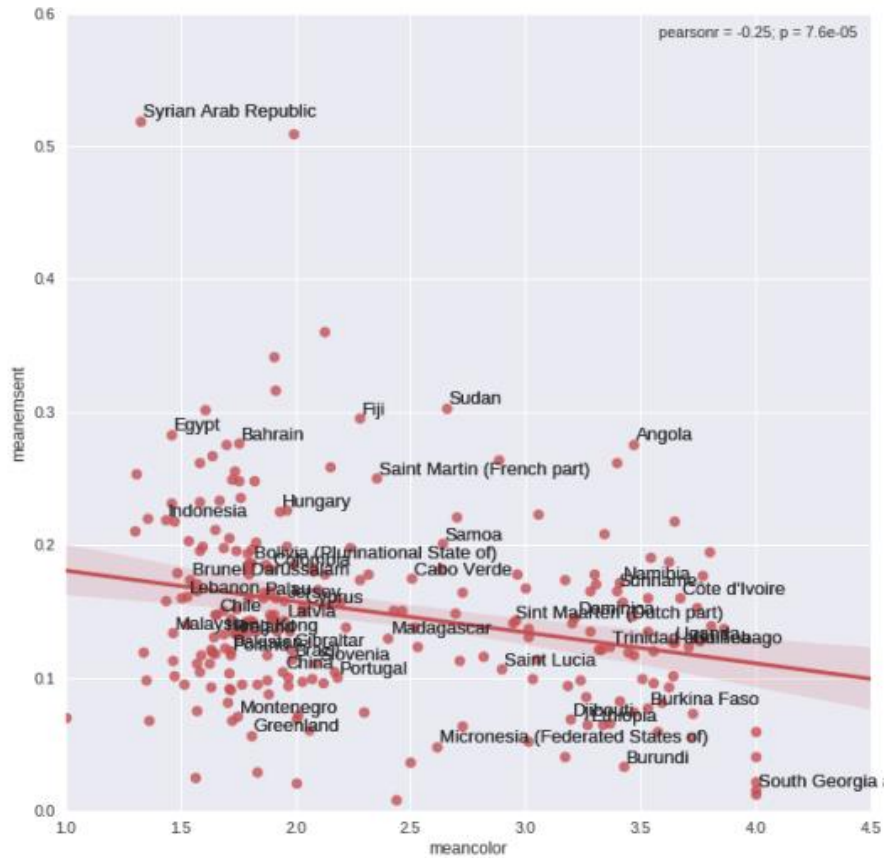

**Fig. 7.** Correlation between Mean Tweet Sentiment and Mean Tweet Skin Tone for 237 Countries/Territories (Shaded Area = 95% Confidence Interval)

To mitigate the effects small sample size (e.g. for countries in which only one or a few users contributed most or all of the skin-tone emoji), the model was refitted for the 50 countries/territories with the highest number of tweets (Fig. 8), with the result that the negative relationship between mean tweet sentiment and mean tweet skin tone strengthened to $r = -0.28$ ($df = 48$, $p = 0.051$).

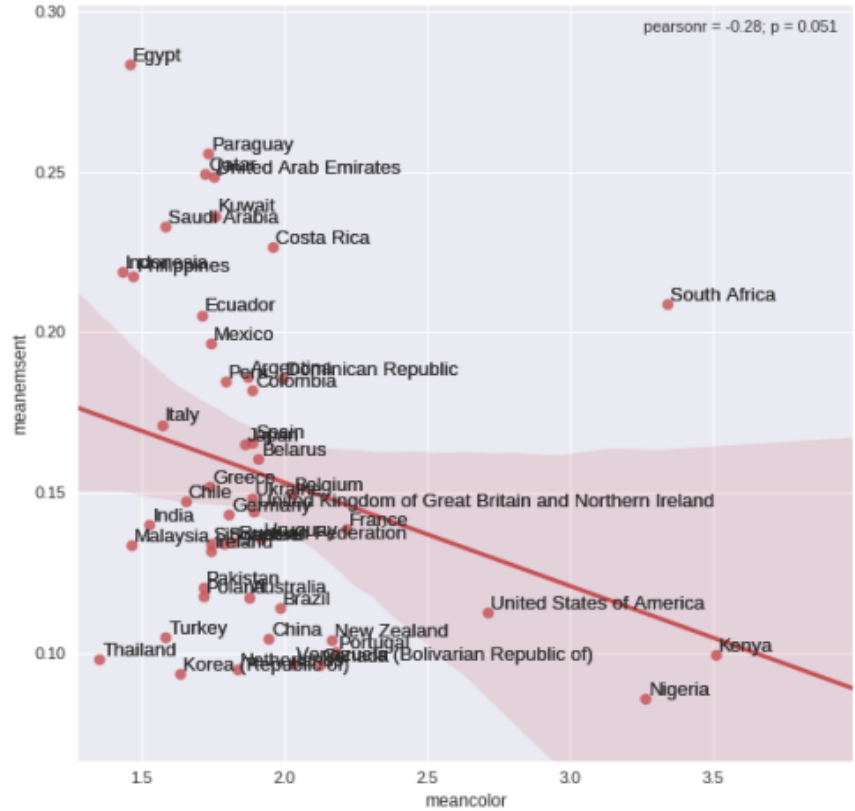

**Fig. 8.** Correlation between Mean Tweet Sentiment and Mean Tweet Skin Tone for 50 Countries with the Largest Number of Tweets (Shaded Area = 95% Confidence Interval)

The countries with lighter mean skin tone values, such as Egypt, Paraguay, Qatar, Indonesia, or the UAE, have higher mean sentiment scores. European countries, with mean skin tone values ranging from approximately 1.7 to 2.25, have middling sentiment values that fall within the 95% confidence interval. The countries with high mean skin tone values, such as the United States, Nigeria, or Kenya, have lower mean sentiment values.

## 5    Summary and Discussion

Since their introduction into the Unicode scheme in 2015, skin tone emoji have become a widely used resource on the Twitter platform. Their global distribution, semantic properties, and patterning with tweet sentiment were investigated in a large corpus of tweets containing geographical metadata by using word embeddings and an emoji sentiment lexicon [3]. Some caveats apply when using the Kralj-Novak et al. lexicon: The number of emoji in the Unicode Standard has increased considerably since creation of the resource. So, for example, an emoji such as 🖕 (U+1F595 REVERSED HAND WITH MIDDLE FINGER EXTENDED), introduced with Emoji 7.0 in 2014, does not have a value in the classification scheme, although it is likely to be used to mark negative affect. Likewise, emoji introduced since 2015, such as sequences consisting of two or more code points joined together (used to represent e.g. flags or groups of persons), are not in the lexicon, and nor are the skin tone emoji themselves. Nonetheless, the broad coverage of the lexicon, in which most of the emoji which can be paired with skin tone emoji are assigned a sentiment value, makes sentiment inference of tweets containing skin tone emoji feasible. The geographical distribution of skin tone emoji, their semantics, and the correlation of skin tone and sentiment suggest several preliminary interpretations.

The global distribution of skin tone emoji shows that to a certain extent, the characters are being used on Twitter as intended by the originators of the Unicode proposal: to make it possible for people to represent their own skin color in online communication. While darker skin tone emoji are used in Africa and the United States, lighter skin tones are more numerous globally and are more likely to be used in Asia, the Middle East, and parts of Latin America and Europe. For the U.S., the prevalence of darker skin tone emoji may in part be explained by the popularity of Twitter among African-Americans, who are overrepresented on the platform compared to their share of the population [16]. For Europe, darker emoji skin tones may indicate a youthful Twitter user population: in general, younger users are more likely to utilize non-standard linguistic resources such as emoji on CMC [26], and in Europe younger people (presumably including some Twitter users) are more likely to come from immigrant backgrounds. In Asia, the Middle East, and parts of Latin America, lighter skin tone emoji may reflect cultural norms concerning the physical attributes of health and beauty in which skin color can play an important role [27], a fact that has been documented in research into body satisfaction [28], attractiveness ratings [29], or use of skin whitening products, particularly by females [30].

The semantics of skin tone emoji are, in part, manifest in a multidimensional vector space model. Lighter skin tone emoji are more similar in their collocational properties (and hence semantics) to other emoji that can be interpreted as expressing generally positive affect, such as smiling faces and heart symbols, while darker skin tone emoji are more closely associated with symbols that express other affective states, including distress, as well as non-standard word forms. The finding corresponds to that of Ljubešić and Fišer [10], who do not investigate sentiment or emoji skin tone directly, but cluster countries based on their emoji distributions. They note that the two darkest skin-tone values, as well as several emoji depicting unhappy faces, belong to a "fourth world" cluster of mainly African countries.

The association between darker emoji skin tone and (possible) negative affect is also manifest when sentiment and mean emoji skin tone and sentiment are regressed. The negative relationship between sentiment and skin tone is weak when all tweets with skin tone emoji are considered, having a value of *r = -0.09*. Because there are so

14many more skin tone emoji tweets in the data from the United States than from other countries, the value is almost the same as that for the United States data alone. When considered at country- or territorial level, however, the value is more strongly negative at $r = -0.25$, increasing in strength to $r = -0.28$ when only the 50 countries with the most tweets are considered. The negative association between sentiment and emoji skin tone in this data is in accord with the implicit findings of the Ljubešić and Fišer study, and parallels the results of survey-based measures of happiness by country, in which many countries of the developing world, primarily in Africa, report low levels of well-being and happiness [31].

The large amount of data collected in this study makes more specific country-level analyses of skin tone emoji use possible. Considering the fact that language- and geography-based differences in emoji usage and meaning have been found [8], the semantics of skin tone emoji in particular languages, countries, or geographical contexts could be more closely examined using vector spaces. Other future work could include updates and refinements to the emoji sentiment lexicon, as well as the utilization of more sophisticated sentiment models based on machine learning, support vector machines, or neural networks. Parsed data containing skin tone emoji could be analyzed to consider evaluative use of skin tone emoji.

As skin tone emoji continue to gain in popularity worldwide, techniques for measuring and evaluating the ways in which they are used are likely to play a role in NLP tasks pertaining to information extraction in bi- and multilingual contexts. In a broader perspective, the analysis of skin tone emoji use can give insight into on how humans represent themselves on social media, what kinds of attitudes and meanings are associated with skin color, and how language is used to depict the phenotypical diversity of the shared human condition.

**Acknowledgement**

The author thanks Finland's Centre for Scientific Computing (CSC) for providing access to computational and data storage facilities.